\documentclass[reprint,nofootinbib,superscriptaddress,amsmath,amssymb,aps,prd]{revtex4-1}
\usepackage{amsmath,mathrsfs,amsbsy,color,graphicx,bm,amsthm,amsfonts}
\usepackage{bbm}
\usepackage{times}
\usepackage{units}
\usepackage{dcolumn}
\usepackage{graphicx}
\usepackage{epsfig}
\usepackage{epstopdf}
\usepackage[colorlinks,linkcolor=blue,anchorcolor=green,citecolor=blue,CJKbookmarks=True]{hyperref}
\DeclareMathSymbol{\shortminus}{\mathbin}{AMSa}{"39}
\usepackage{mathrsfs}
\usepackage{braket}
\usepackage{amssymb}
\usepackage{txfonts}
\usepackage{float}
\usepackage{enumitem}
\usepackage{multirow}

\begin{document}
\title{Open quantum battery in the background of a three-dimensional  rotating  black hole}	
\author{Xiaofang Liu}
\affiliation{Department of Physics, Key Laboratory of Low Dimensional Quantum Structures and Quantum Control of Ministry of Education, and Synergetic Innovation Center for Quantum Effects and Applications, Hunan Normal
		University, Changsha, Hunan 410081, P. R. China}
	
\author{Zehua Tian}
\email{tzh@hznu.edu.cn}
\affiliation{School of Physics, Hangzhou Normal University, Hangzhou, Zhejiang 311121, China}	
	
\author{Jieci Wang}
\email{jcwang@hunnu.edu.cn}\affiliation{Department of Physics, Key Laboratory of Low Dimensional Quantum Structures and Quantum Control of Ministry of Education, and Synergetic Innovation Center for Quantum Effects and Applications, Hunan Normal
		University, Changsha, Hunan 410081, P. R. China}

\begin{abstract}
We investigate the charging performance of a quantum battery coupled to a scalar field in the background of a three-dimensional rotating  black hole.  We show that for Dirichlet boundary conditions, the black hole rotation enhances the charging performance at finite times when the quantum battery's energy level spacing is smaller than the charging amplitude, whereas it degrades the charging performance in other parameter regimes. Notably, as the black hole approaches extremal rotation, charging performance undergoes significant amplification or suppression, depending on the parameter regime. This indicates that the performance of quantum battery can probe critical properties of black holes. Additionally, regarding the energy flow in quantum battery, it is further demonstrated that the energy extraction from vacuum fluctuations via dissipation, and rotation suppresses the quantum battery's capacity to extract this energy. Our findings not only advance the relativistic dissipation dynamics of quantum battery but also propose a novel method to detect black hole rotation and extremal-state transitions.
\end{abstract}
\pacs{~}
\maketitle

\section{Introduction}
In recent years, the rapid advancement of thermodynamics and quantum information science has intensified research focus on the quantum thermodynamic properties at their intersection \cite{esposito2009nonequilibrium, campisi2011colloquium,landi2021irreversible, pekola2021colloquium}.
 An important example is quantum battery  \cite{allahverdyan2004maximal,alicki2013entanglement,yang2023battery}, a miniaturized device that leverages quantum effects and quantum operations to achieve advantages in energy storage capacity and charging power.
To investigate how  the charging performance of quantum battery can be enhanced by quantum effects, various quantum battery schemes have been studied from both theoretical and experimental perspectives \cite{zhu2023charging,niu2024experimental,quach2022superabsorption,joshi2022experimental}, such as the Sachdev-Ye-Kitaev quantum battery \cite{rossini2020quantum,rosa2020ultra}, the cavity quantum battery \cite{crescente2020ultrafast,wang2024cavity}, and the spin-chain quantum battery \cite{grazi2024controlling,rossini2019many,liu2025scrambling}. 
Most early studies focused on closed quantum battery systems, which consist of a quantum battery and charger isolated from environmental interactions \cite{crescente2020ultrafast,zhang2019powerful}. 
However, real quantum systems always interact with the environment \cite{breuer2002theory}, causing relaxation and decoherence. For quantum battery, these effects can significantly impact the charging performance. Therefore, investigating dissipative charging processes in open quantum systems is of critical importance \cite{barra2019dissipative,zhao2021quantum,bhattacharyya2024nonlinearity,malavazi2024weak,hu2025enhancing,ghosh2025constructive,ahuja2025enhancing}.


On the other hand, as a result of quantum field scattering by spacetime, the geometric and quantum properties of spacetime are typically encoded into the field correlation function, which in turn may influence the dynamical evolution of the detector through its interaction with the external quantum field  \cite{benatti2003environment,benatti2004entanglement,soares2022entanglement,yu2011open,kaplanek2023effective,wu2025entangled,moustos2025quantum,moustos2025surpassing}.
Consequently, the interplay between quantum fields and spacetime geometry has attracted considerable attention, ranging from open quantum system approaches in relativistic settings to broader investigations in quantum gravity and quantum information theory, including studies on field correlations and spacetime structure \cite{tian2013geometric,liu2024entanglement,du2021fisher,tian2015relativistic,tian2023using,wu2023does,Wu:2023spa,Wu:2025euf,Li:2025jlu,Liu:2024soc,Liu:2024iec,liu2025lorentz}, as well as vacuum entanglement harvesting and related phenomena \cite{liu2025harvesting,bueley2022harvesting,liu2025wormhole,yang2025analytic,lopez2025quenched,wang2025harvesting,ng2022little}, among others.
In addition, this detector-field interaction model has also been extensively employed to investigate the interplay between the thermodynamics of quantum systems and the superposition property of spacetime  \cite{henderson2020quantum,tang2025can,foo2022quantum,foo2023quantum}. In addition, a quantum battery  \cite{mukherjee2024enhancement,liu2025dissipation} modeled as an Unruh-DeWitt detector  was examined to investigate how the vacuum fluctuations of a massless scalar field affected the battery's charging performance in the background of a static (2+1)-dimensional Ba\~nados-Teitelboim-Zanelli (BTZ) black hole spacetime \cite{tian2025dissipative}. However, the current understanding of how frame-dragging induced by a rotating black hole influences the performance of a quantum battery remains limited. A black hole possessing nonzero angular momentum alters the structure of the surrounding quantum vacuum, making it qualitatively distinct from that associated with a non-rotating black hole. Studies have indicated that rotation can significantly amplify both the entanglement harvested by Unruh-DeWitt detectors \cite{robbins2021entanglement} and the strength of the weak anti-Hawking effect \cite{robbins2022anti}.

	
Motivated by the above considerations, we study the dissipative dynamics of an open quantum battery in the background of a three-dimensional rotating BTZ black hole.
The quantum battery is modeled as a two-level system coupled to the external massless scalar field. By focusing on dissipative dynamics, the charging performance of the battery is analyzed to investigate the interplay between the thermodynamics of quantum systems and gravitational effects. Specifically, we explore how the rotation of black holes affects the charging dynamics and overall performance of quantum battery compared with the static BTZ case. Our results show that when the energy level spacing is smaller than the charging amplitude, dissipation for Dirichlet boundary conditions in the extremal black hole scenario significantly enhances both energy storage capacity and charging power. Furthermore, an analysis of energy flow demonstrates that increased black hole rotation speeds reduce the proportion of energy extracted from vacuum fluctuations.

The organization of the paper is as follows. In Sec. \ref{Sec.2}, we introduce the open quantum battery model. In Sec. \ref{Sec.3}, we introduce the rotating BTZ black hole and derive the charging dynamics of an open quantum battery in this background. Based on the derived dynamics expression, we analyze the effect of dissipation on the average energy and charging power of the quantum battery. Finally, our conclusions are given in Sec. \ref{Sec.4}.

\section{Dissipation dynamics of open quantum battery}\label{Sec.2}
The Hamiltonian of the  quantum battery and quantum field   reads
	\begin{equation}
		H=H_{\mathrm{B}}+H_{\mathrm{C}}+H_{\mathrm{FI}},
	\end{equation}
where $ H_{\mathrm{B}}=\frac{\Delta}{2}\sigma_{z} $ and $ H_{\mathrm{C}} =-\frac{A}{2}\sigma_{x}$ denote the free Hamiltonian of the quantum batttery and the charger, respectively.	
$ \Delta $ is the energy level spacing between the ground state $\left|g\right\rangle$ and the excited state $\left|e\right\rangle$ of the quantum battery, $\sigma_{m}(m=x,y,z)$ represents the $m\mathrm{-th}$ Pauli matrix, and $A$ denotes an external classical field coupled with the $\sigma_{x}$ component of the quantum battery \cite{binder2015quantacell,campaioli2017enhancing,ferraro2018high}.
Meanwhile, $ H_{\mathrm{FI}} = H_{\mathrm{F}} + H_{\mathrm{I}}$ contains two terms: $H_{\mathrm{F}}=\int d^3\mathbf{k}\omega_\mathbf{k}c_\mathbf{k}^\dagger c_\mathbf{k}\frac{dt}{d\tau}$ denotes the free Hamiltonian of the scalar field, and 
$H_{\mathrm{I}}={\lambda}(\sin\theta\sigma_{z}+\cos\theta\sigma_{x})\otimes\phi[x(\tau)]$ denotes the interaction between the quantum battery and the field.
Here, $c_\mathbf{k}^{\dagger}$($c_\mathbf{k}$) is the creation (annihilation) operator of the field mode $\mathbf{k}$ with frequency $ \omega_\mathbf{k} $, $\tau$ denotes the proper time of the quantum battery, $\lambda$ represents the coupling strength between the quantum battery and the field, and $\theta$ characterizes the weight of the quantum battery-field coupling that along the longitudinal $(z)$ and transverse $(x)$ directions. Note that this interaction can capture both decoherence $(\theta=0)$ and pure dephasing $(\theta=\pi/2)$ process.
It is the interaction with the field that will lead to the dissipation of the quantum battery, as shown below.

We assume that the initial state of the quantum battery and field is prepared at $\rho_{\mathrm{tot}}(0)=\rho_{\mathrm{B}}(0)\otimes|0\rangle\langle0|$, where $\rho_{\mathrm{B}}(0)=\frac{1}{2}(\mathrm{I}+\vec{\mathrm{\textbf{b}}}(0)\cdot\vec{\sigma})$, and $\left|0\right\rangle$ represents the vacuum state of the field. At $\tau = 0^{+} $, we switch on the charger, then the quantum battery begins to be charged. To learn the dynamics of the quantum battery conveniently,
we employ a rotation operation in Eq.(\ref{A2}) on the spin space of the quantum battery. Correspondingly, the rotated time-dependent reduced density matrix of quantum battery reads
\begin{equation}\label{eq12}
\tilde{\rho}_{\mathrm{B}}(\tau)=\frac{1}{2}(\mathrm{I}+\vec{\mathbf{a}}(\tau)\cdot\vec{\sigma}).
\end{equation}
At $\tau=0$, the rotated initial reduced density matrix of the quantum battery is given by 
 $\tilde{\rho}_{\mathrm{B}}(0)=\frac{1}{2}(\mathrm{I}+\vec{{\textbf{a}}}(0)\cdot\vec{\sigma})$,
with $ a_{1}(0)=b_{1}(0)\cos\Theta+b_{3}(0)\sin\Theta$, $ a_{2}(0)=b_{2}(0)$,  and $ a_{3}(0)=b_3(0)\cos\Theta-b_1(0)\sin\Theta$, and $\Theta=\mathrm{Arccos}\big({\Delta}/\Omega\big),$ and $ \Omega={\sqrt{\Delta^{2}+\mathrm{A}^{2}}} $.

For the assumption of  weak-coupling limit between the quantum battery and the field, we apply the Born approximation and the Markov approximation \cite{breuer2002theory} to derive the equation of motion of the quantum battery. After that, we obtain the reduced density matrix $\tilde{\rho}_{\mathrm{B}}(\tau)$ yields the Lindblad master equation \cite{manzano2020short,breuer2002theory}
\begin{eqnarray}\label{eq14}
\nonumber
    	\frac{\partial\tilde{\rho}_{\mathrm{B}}(\tau)}{\partial \tau}&=&-i\big[\tilde{H}_{\mathrm{eff}},\tilde{\rho}_{\mathrm{B}}(\tau)\big]+\sum_{i=\pm,z}\bigg[\mathcal{L}_{i}\tilde{\rho}_\mathrm{B}(\tau)\mathcal{L}_{i}^{\dagger}-\frac{1}{2}\big(\mathcal{L}_{i}^{\dagger}\mathcal{L}_{i}\tilde{\rho}_\mathrm{B}(\tau)
\\
&&+\tilde{\rho}_\mathrm{B}(\tau)\mathcal{L}_{i}^{\dagger}\mathcal{L}_{i}\big)\bigg],
    \end{eqnarray}
where $\tilde{H}_{\mathrm{eff}}=\frac{1}{2}\Omega\sigma_{z}=\frac{1}{2}\{\sqrt{\Delta^{2}+A^{2}}+\mathrm{Im}(\Gamma_{+}+\Gamma_{-})\}\sigma_{z} $ is the effective Hamiltonian, and $ \Omega $ denotes the effective energy level-spacing of the quantum battery, where the correction term $\mathrm{Im}(\Gamma_{+}+\Gamma_{-})$ is the Lamb shift. The Lamb shift term is far smaller than $\sqrt{\Delta^{2}+A^{2}}$, and usually can be neglected. In addition, it is found that $\mathcal{L}_{+}=\sqrt{\gamma_{+}}\sigma_{+},\mathcal{L}_{-}=\sqrt{\gamma_{-}}\sigma_{-}$ and $\mathcal{L}_{z}=\sqrt{\gamma_{z}}\sigma_{z}$, with $\sigma_{\pm}={\frac{1}{2}}(\sigma_{x}\pm i\sigma_{y})$ and 
\begin{eqnarray}\label{eq16}
\nonumber
    	 \gamma_{\pm}&=&\lambda^{2}\cos^{2}(\theta-\Theta)\int_{-\infty}^{+\infty}d\Delta \tau e^{\mp i\Omega\Delta \tau}W(x(\tau), x(\tau^\prime)), 
	 \\
    	 \gamma_{z}&=&\lambda^{2}\sin^{2}(\theta-\Theta)\int_{-\infty}^{+\infty}d\Delta \tau W(x(\tau), x(\tau^\prime)),
\end{eqnarray}
where $ \Delta \tau=\tau-\tau^\prime$, and $W(x(\tau), x(\tau^\prime))$ denotes the Wightman function.

Substituting Eq. \eqref{eq12} into Eq. \eqref{eq14}, one can solve this mater equation analytically. The corresponding time-dependent 
state parameters in Eq. \eqref{eq12} are given by 
 \begin{eqnarray}\label{eq17}
 \nonumber
 a_{1}(\tau)&=&a_{1}(0)\cos(\Omega \tau)e^{-\frac{1}{2}(\gamma_{+}+\gamma_{-}+4 \gamma_{z})\tau}-a_{2}(0)\sin(\Omega \tau)
 \\          \nonumber
 &&\times e^{-\frac{1}{2}(\gamma_{+}+\gamma_{-}+4\gamma_{z})\tau}, \\         \nonumber
 a_{2}(\tau)&=&a_{1}(0)\sin(\Omega \tau)e^{-\frac{1}{2}(\gamma_{+}+\gamma_{-}+4\gamma_{z})\tau}+a_{2}(0)\cos(\Omega \tau)
 \\        \nonumber
 &&\times e^{-\frac{1}{2}(\gamma_{+}+\gamma_{-}+4\gamma_{z})\tau}, \\
 a_{3}(\tau)&=&a_{3}(0)e^{-(\gamma_{+}+\gamma_{-})\tau}+\frac{\gamma_{+}-\gamma_{-}}{\gamma_{+}+\gamma_{-}}(1-e^{-(\gamma_{+}+\gamma_{-})\tau}).
\end{eqnarray}
To obtain the time-dependent state $\rho_{\mathrm{B}}(\tau)$, one can rotate the time-dependent state $\tilde{\rho}_\text{B}(\tau)$, i.e., 
$\rho_{\mathrm{B}}(\tau)=\mathcal{R}^\dagger\tilde{\rho}_\text{B}(\tau)\mathcal{R}$. Note that $\frac{1}{2}(\gamma_++\gamma_-+4\gamma_z)$ denotes the dephasing rate, and $\gamma_++\gamma_-$ represents the incoherent relaxation rate, shown in Eq. \eqref{eq17}, which are responsible for the dissipation 
and decoherence. Furthermore, the parameters $\gamma_\pm, \gamma_z$, as shown Eq. \eqref{eq16}, depend on the field Wightman function, which is related to the spacetime background and the detector's world line. 

 \section{Quantum battery in rotating BTZ black hole}\label{Sec.3}
We investigate a quantum battery coupled to a conformal massless scalar field in the vicinity of a rotating BTZ black hole, with particular attention to how the black hole's rotation influences the charging performance of the quantum battery through dissipation effects. The line element for the rotating BTZ black hole is expressed as follows \cite{banados1992black}
    \begin{eqnarray}
    	ds^2=-h(r)dt^2+\frac{1}{h(r)}dr^2+r^2\left[N_\varphi(r)dt+d\varphi\right]^2,\label{eq20}
    \end{eqnarray}
where $ h(r)=-M+\frac{r^2}{\ell^2}+\frac{J^2}{4r^2}, $ $N_{\varphi}(r)=-\frac{J}{2r^{2}}$, with $M$ being the mass of the black hole, 
$J$ being its angular momentum, and $ \ell $ being the AdS length.  Note that the metric in Eq.(\ref{eq20}) is a vacuum solution of Einstein field equations with a negative cosmological constant. The inner and outer horizons (where $h(r)=0$) of the black hole read
\begin{eqnarray}\label{ridus}
    	r_{\pm}^{2}=\frac{1}{2}\left(M\ell^{2}\pm\sqrt{M^{2}\ell^{4}-\ell^{2}J^{2}}\right).
\end{eqnarray}
Notably, \( |J| \leq M\ell \), and in the below analysis, we consider $ J \in [ 0,1 ) $. When $ J/M\ell $ = 0, it indicates a static BTZ black hole. An increase in $ J/M\ell $ corresponds to an increase in the rotational velocity of the black hole. The condition \( r_+ = r_- \) corresponds to the extremal condition.

To derive the charging dynamics of the quantum battery in the rotating BTZ spacetime, one first needs to known the Wightman function of the scalar field. In this paper, we assume that the scalar field is initially in the Hartle-Hawking vacuum. 
Since the BTZ spacetime is locally equivalent to the $\mathrm{AdS}_{3}$ spacetime, 
the Wightman function for the  conformal massless scalar field in the rotating BTZ can be represent by  \cite{carlip19952+,lifschytz1994scalar}
    \begin{eqnarray}\label{WFunction}
    \nonumber
  W_{\mathrm{BTZ}}(x,x^{\prime})&=&\sum_{n=-\infty}^{\infty}\eta^{n}W_{\mathrm{AdS_{3}}}(x,\Gamma^{n}x^{\prime})  
 =\frac{1}{4\pi\sqrt{2}\ell}\sum_{n=-\infty}^{\infty}\eta^{n}
 \\
 &&\times\bigg[\frac{1}{\sqrt{\sigma_{\epsilon}(x,\Gamma^{n}x^{\prime})}} -\frac{\zeta}{\sqrt{\sigma_{\epsilon}(x,\Gamma^{n}x^{\prime})+2}}\bigg],
    \end{eqnarray}
where $\eta=\pm1$ denotes the untwisted/twisted scalar field case, and $\Gamma :(t, r, \varphi)\rightarrow(t, r, \varphi+2\pi)$ represents the action of the identification. The parameter $\zeta\in\{-1,0,1\}$ represents the field's boundary conditions at spatial infinity \cite{avis1978quantum}: Neumann, transparent, and Dirichlet, respectively. Besides, we have 
 \begin{eqnarray}
 \nonumber    		
\sigma_\epsilon(x,\Gamma^nx^{\prime})^2&=& -1+\frac{\cosh\bigg[\frac{r_{+}}{\ell}(\Delta\phi-2\pi n)-\frac{r_{-}}{\ell^{2}}(t-t^{\prime})\bigg]}
{\big[\alpha(r)\alpha(r^{\prime})\big]^{-1/2}}
 \\
&&-\frac{\cosh\bigg[\frac{r_{+}}{\ell^{2}}(t-t^{\prime})-\frac{r_{-}}{\ell}(\Delta\phi-2\pi n)\bigg]}{\big[(\alpha(r)-1)(\alpha(r^{\prime})-1)\big]^{-1/2}},
\end{eqnarray}
where $\alpha(r)=\frac{r^2-r_-^2}{r_+^2-r_-^2}$, and $\Delta\phi=\phi-\phi^{\prime}$.     
    
 Here we only consider the untwisted scalar field case. We assume that our quantum battery is co-rotating with the black hole, thus the corresponding world line yields
 \begin{eqnarray}\label{WorldL}
  \nonumber
    	t&=&\frac{\ell r_+\tau}{\sqrt{r^2-r_+^2}\sqrt{r_+^2-r_-^2}},\\
	  \nonumber
	r&=&R,     \\          
    	\phi&=&\frac{r_{-}\tau}{\sqrt{r^{2}-r_{+}^{2}}\sqrt{r_{+}^{2}-r_{-}^{2}}}.
    \end{eqnarray}
Substituting the Wightman function in Eq. \eqref{WFunction}, together with the quantum battery's world line in Eq. \eqref{WorldL}, into Eq. \eqref{eq16}, one can derive    
 \begin{widetext}
 \begin{eqnarray}\label{eq30}
 \nonumber
\gamma_{\pm}&=&\frac{\lambda^2 \cos^2(\theta-\Theta)}{4}\bigg[1-\tanh\left(\pm\frac{\Omega}{2T}\right)\bigg]\sum_{n=-\infty}^{\infty}e^{\pm\frac{in\Omega r_-}{\ell T}}\bigg[P_{\pm\frac{i\Omega}{2\pi T}-\frac{1}{2}}\big(\cosh\alpha_n^-\big)-\zeta P_{\pm\frac{i\Omega}{2\pi T}-\frac{1}{2}}\big(\cosh\alpha_n^+\big)\bigg], \\
\gamma_{z}&=&\frac{\lambda^2 \sin^2(\theta-\Theta)}{4}\sum_{n=-\infty}^{\infty}\bigg[P_{-\frac{1}{2}}\big(\cosh\alpha_n^-\big)-\zeta P_{-\frac{1}{2}}\big(\cosh\alpha_n^+\big)\bigg],
\end{eqnarray}
 \end{widetext}
where $ T $ is the local KMS temperature, which is the temperature experienced by the quantum battery, 
satisfying $T=T_{H}/\gamma$, with $\gamma=\frac{\sqrt{R^{2}-r_{+}^{2}}\sqrt{r_{+}^{2}-r_{-}^{2}}}{lr_{+}}$ and $ T_H=\frac{1}{2\pi\ell^2}\left(\frac{r_+^2-r_-^2}{r_+}\right)$. Besides, $ P_{v}\left(x\right) $ is the associated Legendre function of the first kind, satisfying $ P_{-\frac{1}{2}+ i\lambda} = P_{-\frac{1}{2}- i\lambda}$, and the
definitions 
\begin{equation}
\cosh(\alpha_n^\pm) = {\pm}4\ell^2 \pi^2 T^2 + (1 + 4\ell^2 \pi^2 T^2) \cosh\left(\frac{2\pi n r_{+}}{\ell}\right)
\end{equation}
have been used.    
     
With the above specific parameters, $\gamma_\pm$ and $\gamma_z$, we can analyze how the properties of spacetime influence the dynamics of the quantum battery, thereby affecting its relevant charging performance. Specifically, we will examine the charging performance of the quantum battery in terms of its figures of merit, namely, the average energy stored in the quantum battery during the charging protocol and the average charging power that characterizes the charging speed of the protocol \cite{binder2015quantacell}.

\subsection{Dynamics of the average energy}

In this subsection, we investigate how the rotation property of the rotating BTZ black hole affects  the average energy stored 
in the quantum battery during the charging protocol,
through dissipation determined by $\gamma_\pm$ and $\gamma_z$ shown in Eq. \eqref{eq30}.  In doing so, we assume the quantum battery is initially prepared in the ground state $ |g\rangle $, i.e., $ b_1(0) = 0, b_2(0) = 0, b_3(0) = -1$, and consider the decoherence coupling case ($\theta = 0$ shown in $ H_I $). We can obtain the average energy of the quantum battery and the charger \cite{yang2023battery}
\begin{eqnarray}\label{eq32}
\nonumber
\langle E_{\mathrm{B}}(\tau) \rangle &= &\frac{\Delta}{2} \bigg\{ 
		1 - \frac{\Delta^2}{\Omega^2} e^{-\mathcal{K}'\tau} 
		- \frac{\Delta}{\Omega} \left( \frac{e^{\Omega/T} - 1}{e^{\Omega/T} + 1} \right) (1 - e^{-\mathcal{K}'\tau}) \\
		& &\quad\;\;\ - \frac{A^2}{\Omega^2} \cos(\Omega \tau) e^{-\mathcal{K}\tau} 
		\bigg\},
\end{eqnarray}
\begin{eqnarray}\label{eq33}
\nonumber
\langle E_{\mathrm{C}}(\tau) \rangle 
		&=& \frac{A^2}{2 \Omega} \Bigg\{ 
	 - \frac{\Delta}{\Omega} e^{-\mathcal{K}'\tau} 
		- \left( \frac{e^{\Omega/T} - 1}{e^{\Omega/T} + 1} \right) (1 - e^{-\mathcal{K}'\tau}) \\
		&& \quad\quad\;\ + \frac{\Delta}{\Omega} \cos(\Omega \tau) e^{-\mathcal{K}\tau} 
		\Bigg\},
\end{eqnarray}
where
   \begin{eqnarray}\label{eq34}
   \nonumber
   \mathcal{K}^\prime &=& \frac{\lambda^2 \Delta^2}{2\Omega^2}    \sum_{n=-\infty}^{\infty}e^{\frac{i\Omega nr_-}{\ell T}} \bigg[P_{\frac{i\Omega}{2\pi T}-\frac{1}{2}}\big(\cosh\alpha_n^-\big)-\zeta P_{\frac{i\Omega}{2\pi T}-\frac{1}{2}}\big(\cosh\alpha_n^+\big)\bigg],  \\
    	 \mathcal{K} &=& \frac{1}{2} \mathcal{K}^\prime + \frac{\lambda^2 A^2}{2\Omega^2} \sum_{n=-\infty}^{\infty} \bigg[ P_{-\frac{1}{2}} (\cosh\alpha_n^-) - \zeta P_{-\frac{1}{2}} (\cosh\alpha_n^+) \bigg].\label{infiniteseries}
    \end{eqnarray}
Note that dissipation effects are reflected in the incoherent relaxation rate  $\mathcal{K}^\prime$ and the dephasing rate $\mathcal{K} $. Specifically, Eq.~(\ref{infiniteseries}) involves an infinite series \(n\); to ensure the numerical accuracy of the results presented, we performed numerical convergence tests. The results indicate that the numerical values have converged and become stable when the summation is truncated at $n=5$. Including additional terms beyond $n=5$ does not affect the reported results but increases computational cost. Therefore, in the following section, we set the truncation order $n=5$ when calculating the battery's average energy.


The absence of dissipation between the quantum battery and the field is also a scenario we are concerned about, which corresponds to the coupling constant $ \lambda $ = 0 in $ H_I $, when the closed quantum battery is driven by a static protocol and all its all energy is supplied by the charger. In this case, the energy stored in the quantum battery is
\begin{eqnarray}\label{emax}
	\langle \, E_{\mathrm{B}}(\tau) \rangle = \frac{\Delta}{2} \frac{A^{2}}{\Omega^{2}} \left[ 1 - \cos(\Omega \tau) \right].
\end{eqnarray}
From Eq. (\ref{emax}), we know that the energy of the quantum battery varies periodically and that the energy of the closed quantum battery reaches its maximum when $ \Omega \tau = \pi n\ (n > 0 \text{ odd numbers}) $, i.e., $ \langle E_{\mathrm{B}}(\tau) \rangle_{\mathrm{static}} = \Delta \frac{A^2}{\Omega^2} $. And when $ A \gg \Delta $, the stored energy $ \langle E_{\mathrm{B}}(\tau) \rangle_{\mathrm{static}} $ approaches its maximum value.

\begin{figure}[h]
	\centering
	\includegraphics[scale=0.75]{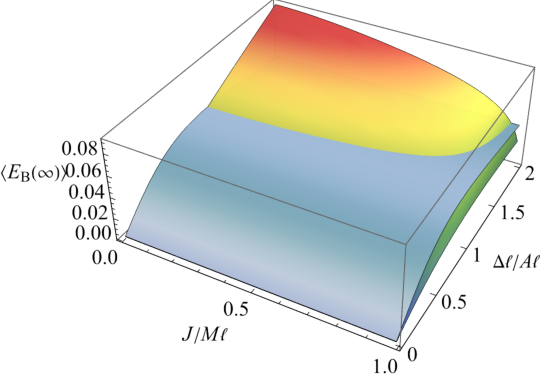}
	\caption{ The average energy \( \langle E_{\mathrm{B}}(\infty) \rangle \) stored in the quantum battery, given by Eq. (\ref{eq35}), parameterized by the angular momentum \( J/M\ell \). The gray surface indicates the maximum average energy $ \langle E_{\mathrm{B}}(\tau) \rangle_{\mathrm{static}} $ stored by the closed quantum battery. The parameters are: $ \ell = 1 $, $ M = 1 $, $ A \ell = 0.1 $ and $ R = 1.01 $.
	}
	\label{fig2}
\end{figure}

In addition, there is another special scenario in which the quantum battery interacts with the field in the infinite time limit, i.e., $ \tau \to \infty $. In this limit, the average energy in Eq. (\ref{eq32}) stored in the quantum battery is expressed as
\begin{equation}\label{eq35}
	\langle \, E_{\mathrm{B}}(\infty) \rangle = \frac{\Delta}{2} \biggl\{ 1 - \frac{\Delta}{\Omega} \biggl( \frac{e^{\Omega/T}-1}{e^{\Omega/T}+1} \biggr) \biggr\}.
\end{equation}
Eq.~(\ref{eq35}) agrees with the corresponding expression in the static BTZ case \cite{tian2025dissipative}. This agreement arises because, in the long-interaction limit, the average energy given in Eq.~(\ref{eq32}) is no longer affected by the exponential decay term in Eq. (\ref{eq34}).
This also  suggests the boundary conditions have no discernible effect on the energy at this stage.
Instead, the average energy is solely determined by local KMS temperature $ T $ of rotating BTZ black hole and relative weighting between the quantum battery's energy level spacing $\Delta \ell$ and the charger's amplitude $A \ell$. Since the redshift factor $\gamma$ and the Hawking temperature $ T_H $, which together determine the local temperature $T$, indirectly depend on the black hole's angular momentum $J/Ml$ (see Eq.~(\ref{ridus})), we plot the average energy \( \langle E_{\mathrm{B}}(\infty) \rangle \) as a function of $J/Ml$ and the ratio $\Delta \ell/A \ell$ in Fig.~\ref{fig2} to investigate the impact of rotation on quantum battery dynamics. 

 \noindent

The figure shows that, in the absence of dissipation (i.e., the quantum battery is not coupled to the vacuum field), the maximum energy $ \langle E_{\mathrm{B}}(\tau) \rangle_{\mathrm{static}} $ stored in the closed quantum battery depends only on the energy level spacing \(\Delta \ell\) and the charging amplitude \(A \ell\).
And no energy is stored in the battery when the ratio of  \(\Delta \ell\)/\(A \ell\) = 0.
The reason is that, even when the battery is charged, the energy level spacing \(\Delta \ell\) determines its storage capacity. When \(\Delta \ell\)=0, the battery has zero storage capacity.
As the \(\Delta \ell\)/\(A \ell\) increase, the maximum energy $ \langle E_{\mathrm{B}}(\tau) \rangle_{\mathrm{static}} $ stored in the closed quantum battery first increases sharply, reaches a peak when the ratio of  \(\Delta \ell\)/\(A \ell\) is near one, and then decreases slowly. 
This implies that the charging amplitude \(A \ell\) is fixed, the quantum battery stores its maximum energy when $\Delta \ell = A \ell$, which equals $A/2$.

We find that in the long-time limit quantum battery interacting with the vacuum field may on average store more energy than the closed quantum battery. This behavior occurs when both \(\Delta \ell > A \ell\) and the angular momentum \(J/M\ell\) $ < $ 0.95. At this point, the energy stored in the quantum battery will originate not only from the charger but also from vacuum fluctuations via dissipative processes. Specifically, we find that when the angular momentum $J/M\ell=0$, i.e., for a non-rotating black hole, the quantum battery extracts the maximum possible energy from vacuum fluctuations. However, as the angular momentum \(J/Ml\) increases (that is, as the black hole's rotational speed rises) the energy \( \langle E_{\mathrm{B}}(\infty) \rangle \) gradually decreases. This implies that, in the absence of boundary conditions, the nonzero angular momentum modifies the structure of the vacuum field around the black hole, thereby reducing the quantum battery's ability to extract energy from vacuum fluctuations. Moreover, the black hole's rotation further amplifies this effect.

 \begin{figure}[h]
 	\centering
 	\includegraphics[scale=0.43]{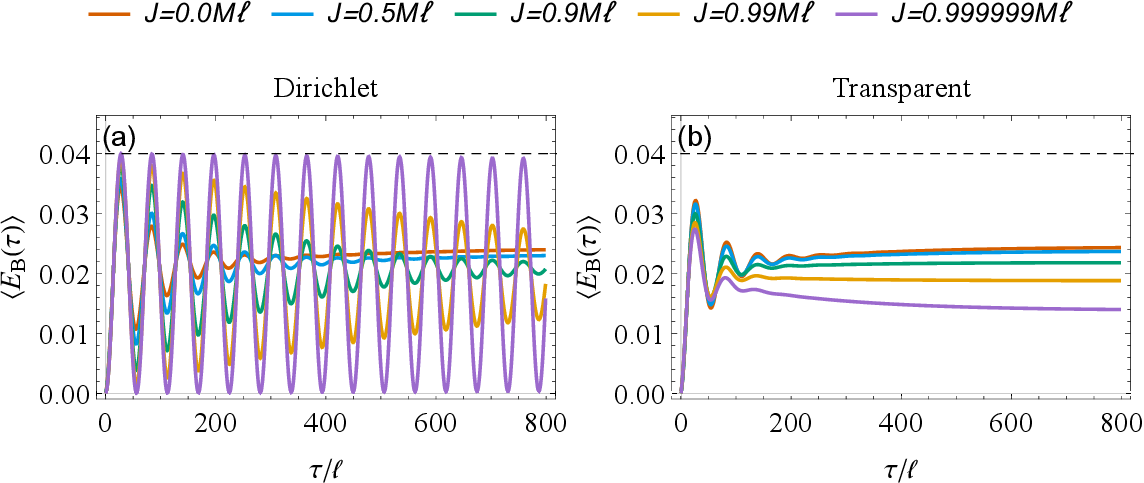}
 	\includegraphics[scale=0.426]{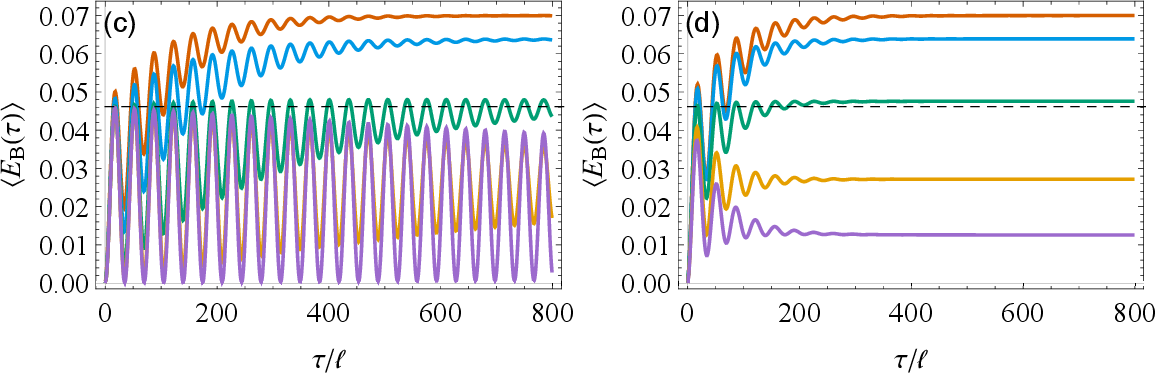}
 	\caption{ Evolution of the average energy $ \langle E_{\mathrm{B}}(\tau) \rangle $  stored in the quantum battery over time. In the first row take $ \Delta \ell $ = 0.05, $ A \ell $ = 0.1. In the second row take $ \Delta \ell $ = 0.15 and $ A \ell $= 0.1. Here we set the mass $ M = 1 $, the radial coordinates of quantum battery $ R = 1.01 $ and coupling strength $ \lambda\sqrt{\ell} = 0.2 $. The black dotted line indicates the maximum average energy of the quantum battery when the coupling strength $\lambda \sqrt{\ell} = 0$. }
 	\label{fig1}
 \end{figure}\noindent	

Fig.~\ref{fig1} shows the average energy $\langle E_{\mathrm{B}}(\tau)\rangle$, given by Eq.~\eqref{eq32}, stored in the quantum battery as a function of the evolution time $\tau/\ell$.
We plot the maximum average energy stored in the closed quantum battery driven by a static driving protocol (corresponding to $\lambda\sqrt{\ell} = 0$ case in the interacting Hamiltonian $ H_I $ as a reference limit (see black dotted curves), which is 
given by $\langle E_{\mathrm{B}}(\tau) \rangle_{\text{static}} = \Delta \frac{A^2}{\Omega^2}$.

From Fig.~\ref{fig1}(a), we find that, when the energy level spacing $\Delta\ell$ of the quantum battery is smaller than the charging amplitude $A\ell$, the average energy $ \langle E_{\mathrm{B}}(\tau) \rangle $ does not exceed that stored in a closed quantum battery for Dirichlet boundary conditions. Moreover, the average energy $ \langle E_{\mathrm{B}}(\tau) \rangle $ exhibits damped oscillations over time. 
This is because the energy transferred from the charger to the quantum battery, due to environmental effects, introduces incoherent relaxation rate $ \mathcal{K}^\prime $ and dephasing rate $ \mathcal{K} $, thereby leading to the system's dissipative evolution. Specifically, the vacuum in the background spacetime of a nonrotating ($ J/Ml = 0 $) BTZ black hole exhibits maximal dissipation. For rotating BTZ black holes ($ J/Ml \neq 0 $), the amplitude of the energy oscillations decreases as the angular momentum $ J/Ml $ increases. This indicates that, because of the interaction between the quantum battery and its environment, the black hole's nonzero angular momentum can reduce the incoherent relaxation and dephasing rates in Eq. (\ref{eq34}), thereby suppressing energy dissipation from the dynamics. When the quantum battery is isolated from its environment, its energy exhibits periodic variations, as described by Eq.~(\ref{emax}). However, as shown in Fig. \ref{fig1}(b), for transparent boundary conditions, the rotational effects of the black hole accelerate the energy dissipation from the dynamics. This effect is attributed to the fact that nonzero angular momentum increases rather than decreases the incoherent relaxation rate $ \mathcal{K}^\prime $ and the dephasing rate $ \mathcal{K} $ in Eq. (\ref{eq34}) for this boundary conditions, resulting in greater dissipation.

For $\Delta\ell>A\ell$, we can see from Fig. \ref{fig1}(c)  and Fig. \ref{fig1}(d) that the average energy exhibits a quite different behavior compared with the $\Delta\ell<A\ell$ case shown in the Fig. \ref{fig1}(a)  and Fig. \ref{fig1}(b). 
Notably, the maximum average energy stored in the open quantum battery can exceed that of the closed quantum battery during the charging process. This suggests that the quantum battery can obtain energy not only from the charger but also from vacuum fluctuations via dissipative processes, similar to the static BTZ black hole \cite{tian2025dissipative}.
Notably, in the static BTZ spacetime (i.e., $J/M l=0$), the quantum battery attains its maximum stored energy, indicating maximal extraction of energy from vacuum fluctuations for this conditions. However, as the black hole starts rotating (i.e., as its angular momentum \(J/Ml\) gradually increases), we observe that the energy stored in the open quantum battery decreases. For \(J/Ml < 0.9\), the energy stored in the open quantum battery falls below that of the closed quantum battery. These results indicate that, as the BTZ black hole transitions from a non-rotating (static) state to a rotating state, an increase in the rotation parameter suppresses energy extraction and thus reduces the quantum battery's energy storage capacity. This outcome arises because the black hole's nonzero angular momentum modifies vacuum fluctuations, thereby modulating the quantum battery's energy dissipation rate and affecting the battery's performance.
This applies to both Dirichlet and transparent boundary conditions. 

The behavior of the average energy also depends on the boundary condition of the quantum field at the spatial infinity very much. Compared Fig. \ref{fig1}(a)  with Fig. \ref{fig1}(b), we can find that the boundary condition may induce absolutely opposite behavior of the average energy when varying the angular momentum  for the $\Delta\ell<A\ell$ scenario. In these two boundary condition cases, they have different maximum and asymptotic evolutionary dynamics. 
However, for the $\Delta\ell>A\ell$ case, the transparent boundary condition and the Dirichlet boundary condition lead to the same behavior of the average energy via the charging time by varying the angular momentum $ J/Ml $, while quantitatively different.
Moreover, the greater the angular momentum $J/M\ell$, the more pronounced the effect of the boundary conditions on the dynamical behavior of the average energy. However, although the average energy stored in the quantum battery during finite-time charging exhibits different evolution for different boundary conditions, in the infinite-time limit in Eq.(\ref{eq35}), the stored energy depends only on the rotation of the BTZ black hole and is independent of the boundary conditions.

  \begin{figure}[h]
  	\centering
  	\includegraphics[scale=0.42]{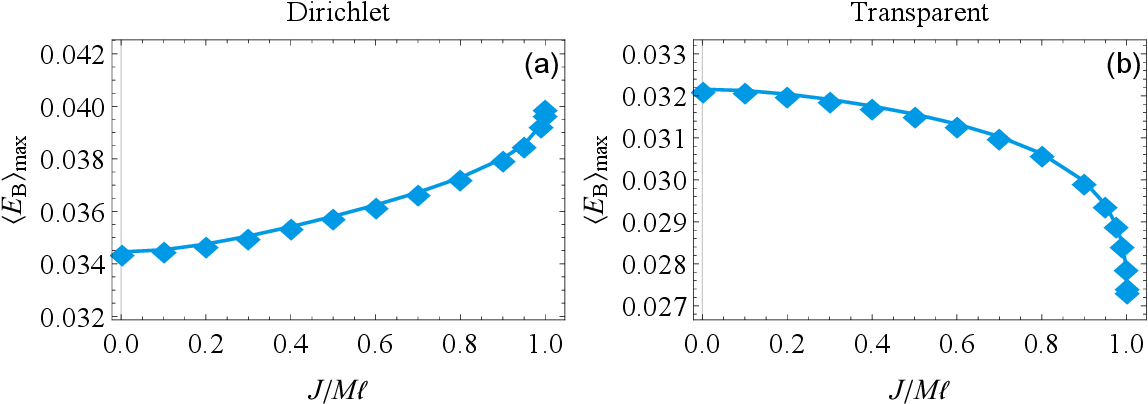}
  	\caption{ Variation of the maximum average energy  $ \langle \, E_{\mathrm{B}} \rangle_{\text{max}} $ of the quantum battery as a function of angular momentum  $ J/M \ell $ when $ \Delta \ell  = 0.05$ and $A \ell = 0.1$.
  		The other parameters are: $ \ell = 1 $, $ M = 1 $ and $ R = 1.01 $.
  	}
  	\label{fig12}
  \end{figure}\noindent

  The maximum energy stored in a quantum battery is also a key metric for evaluating its performance
  \begin{eqnarray}
  	\langle \, E_{\mathrm{B}} \rangle_{\text{max}} = \max_{\tau} [\langle E_{\mathrm{B}}(\tau) \rangle] = \langle E_{\mathrm{B}}(\tau_E)\rangle,
  \end{eqnarray}
  where $\tau_E$ is the time when the average energy reaches its maximum value.
  
 In Fig. \ref{fig12} we plot the maximum average energy $ \langle \, E_{\mathrm{B}} \rangle_{\text{max}} $ stored in the quantum battery as a function of the 
 angular momentum $ J/Ml $ for $\Delta\ell<A\ell$. We observe that the variation in the maximum average energy $ \langle \, E_{\mathrm{B}} \rangle_{\text{max}} $ depends strongly on the boundary conditions. Specifically, with the increase of angular momentum $ J/Ml $, the maximum average energy $ \langle \, E_{\mathrm{B}} \rangle_{\text{max}} $ increases monotonously for the Dirichlet boundary condition case. However, for the transparent boundary case the maximum average energy $ \langle \, E_{\mathrm{B}} \rangle_{\text{max}} $ decreases monotonously when the angular momentum $ J/Ml $ increases.
These two boundary conditions have opposite effects on changes in energy, as in the static black hole scenario \cite{tian2025dissipative}.
 Furthermore, for the extreme case of the rotating BTZ black hole, i.e.,  the angular momentum $ J/Ml $ approaches to $1$, the average energy, for both the Dirichlet and  the transparent boundary cases, is quite sensitive to the variation of the angular momentum $ J/Ml $. This means that as the black hole approaches the extremal limit, even small changes in the rotational speed can trigger abrupt changes in the amount of maximum average energy stored in the quantum battery. This behavior may result from dramatic variations in the vacuum state near the extremal black hole.

 \subsection{Effect of dissipation on charging power}
 In this subsection, we employ the  average charging power 
to characterize the ``charging speed" \cite{binder2015quantacell}, which is defined as the ratio between the average deposited energy $\langle E_{\mathrm{B}}(\tau) \rangle$ and the time $\tau$
required to complete the procedure,
\begin{eqnarray}
	P_\mathrm{B} (\tau) = \frac{\langle E_{\mathrm{B}}(\tau) \rangle} {\tau}.
\end{eqnarray}
Correspondingly, the maximum value of $P_\mathrm{B} (\tau)$ is called the maximal power, given by  
\begin{equation}
	P_{\text{max}} \equiv \max_{\tau} [P_\mathrm{B}(\tau)] \equiv P_\mathrm{B}(\tau_P),
\end{equation}
 where $\tau_P$ is the time when the average charging power reaches its maximum value. We are interested in how the 
 rotation property of the rotating BTZ black hole affects the charing performance of the quantum battery from the perspective
 of the average charging power.
 
 \begin{figure}[h]
	\centering
	\includegraphics[scale=0.55]{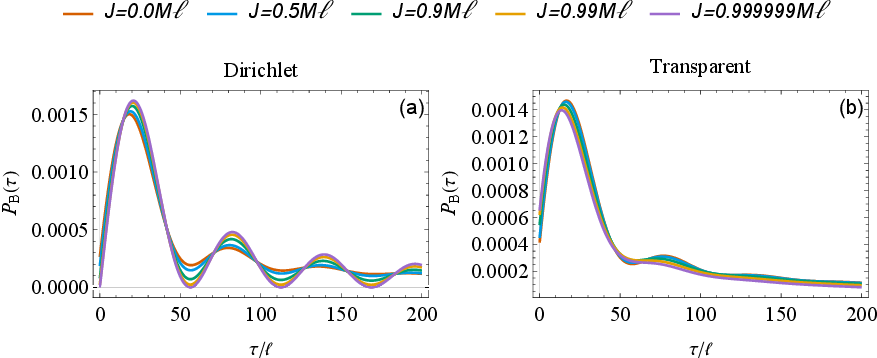}
	\includegraphics[scale=0.55]{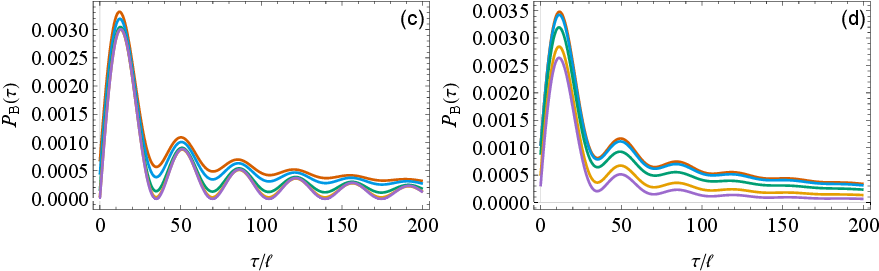}
	\caption{ The average charging power $ P_\mathrm{B}(\tau) $ versus the charging time $ \tau/\ell $ for different values of the angular momentum $ J/M\ell $. (a)-(b): $ \Delta \ell $ = 0.05, $ A \ell $ = 0.1; (b)-(c) $ \Delta \ell $ = 0.15, $ A \ell $ = 0.1.The other parameters are:  $ \ell = 1 $, $ M = 1 $, $ R = 1.01 $ and $ \lambda\sqrt{\ell} = 0.2 $.
	}
	\label{fig3}
\end{figure}	
In Fig. \ref{fig3} we plot the average charing power $ P_\mathrm{B} (\tau) $ as a function of the charging time $ \tau/\ell $ with varying the angular momentum $ J/M\ell $ while fixing other parameters.
In each image, the initial peak of the average power was the largest, and the peak decreased gradually over time. Initially, energy can flow into the battery at the maximum possible rate. However, subsequent oscillations arise because the system is open and interacts with the vacuum field. This interaction induces irreversible dissipation and decoherence. Consequently, vacuum fluctuations act as a noisy environment, causing the system to lose coherence and thereby reducing the charging power. Moreover, we observe in Fig.~\ref{fig3}(a) that, when $\Delta\ell<A\ell$, the charging performance of the quantum battery is minimized in the static BTZ black hole background ($ J/M\ell $ =0) with the Dirichlet boundary condition. However, in the rotating BTZ black hole background ($ J/M\ell  \neq 0$), increasing the angular momentum $J/Ml$ enhances the battery's $P_B(\tau)$. This results imply that black hole rotation enhances quantum battery performance for the considered condition. For other conditions (see Figs.\ref{fig3}~(b)–(d)), increasing the black hole's angular momentum $J/Ml$ reduces the $P_B(\tau)$ of the quantum battery.

\begin{figure}[h]
	\centering
	\includegraphics[scale=0.405]{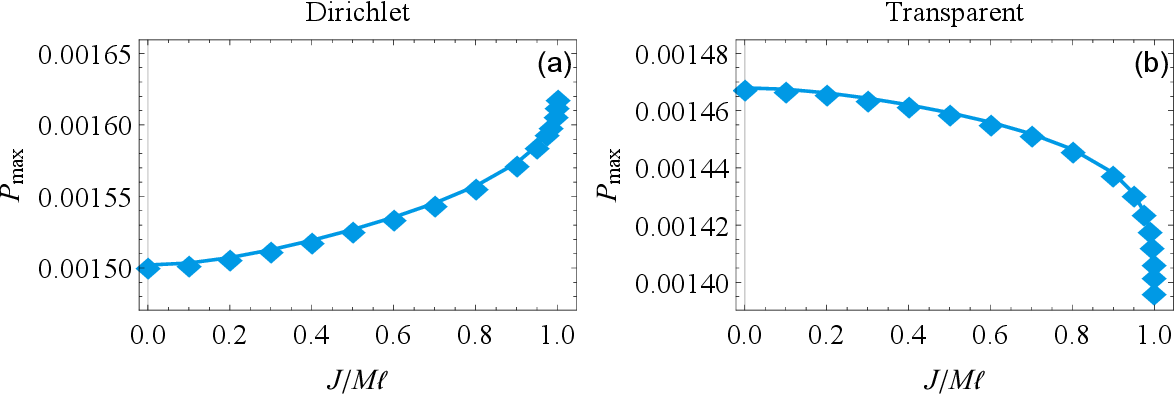}
	\includegraphics[scale=0.405]{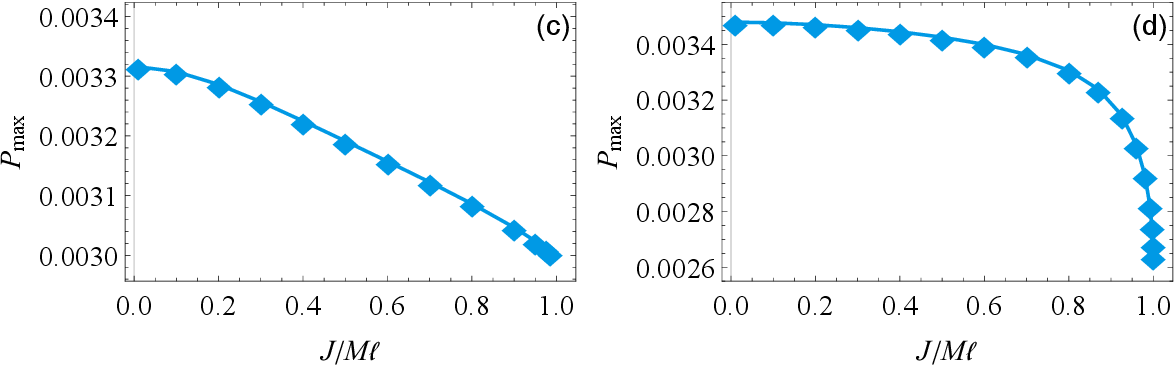}
	\caption{ Variation of the maximum charging power $ P_{\text{max}} $ with the angular momentum $ J/M \ell $. The first row shows $ \Delta \ell $ = 0.05, $ A \ell $ = 0.1; the second row shows $ \Delta \ell $ = 0.15, $ A \ell $ = 0.1. The other parameters are:  $ \ell = 1 $, $ M = 1 $, $ R = 1.01 $ and $ \lambda\sqrt{\ell} = 0.2 $.
	}
	\label{fig5}
\end{figure}\noindent

To obtain a comprehensive understanding of the impact of black hole rotation on the charging process of the quantum battery, we plot the maximum charging power \(P_{\max}\) as a function of the angular momentum \(J/M\ell\) in Fig.~\ref{fig5}. As shown in Fig. \ref{fig5}(a), for Dirichlet boundary conditions and when $\Delta\ell<A\ell$, the maximum charging power \(P_{\max}\) of the quantum battery increases monotonically with the angular momentum \(J/M\ell\). This demonstrates that, for these conditions, optimal charging power is obtained for vacuum fields around rapidly rotating black holes (i.e., black holes with high angular momentum). Notably, unlike Fig.~\ref{fig5}(a), Figs.~\ref{fig5}(c)-(d) show that the maximum charging power \(P_{\max}\) of the quantum battery decreases monotonically as the angular momentum \(J/M l\) increases. The results imply that the quantum battery achieves optimal charging power when coupled to the vacuum field in the background spacetime of a static (non-rotating) BTZ black hole. Moreover, this optimal charging power cannot be increased by increasing the angular momentum \(J/M\ell\). 
 Specifically, when the black hole approaches the extremal limit, where $J/M\ell\to 1$, the charging power varies sharply. This may be attributed to drastic changes in the quantum vacuum as the black hole approaches extremality.

 \subsection{Energy flow in quantum battery}
\begin{figure}[h]
 	\centering
 	\includegraphics[scale=0.413]{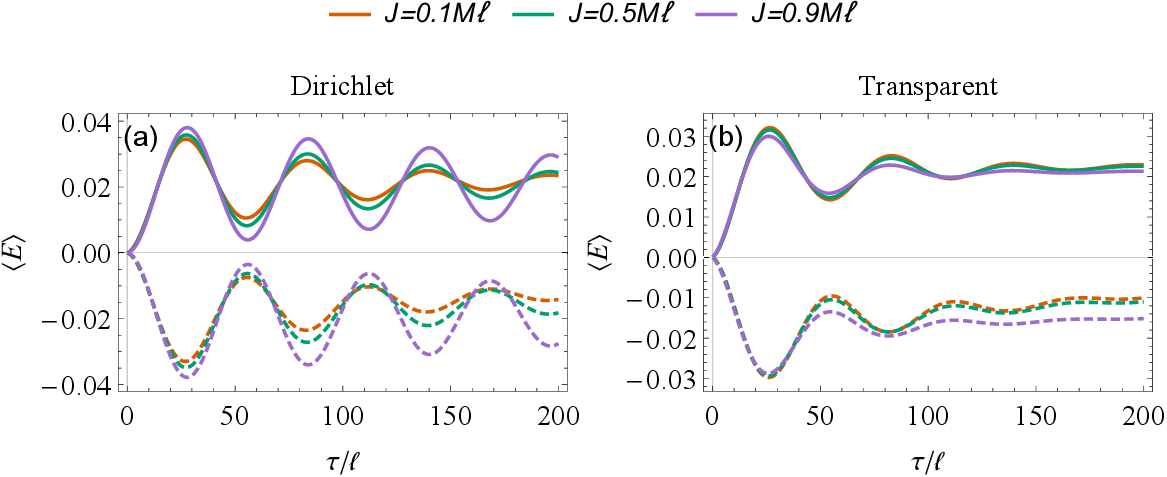}
 	\includegraphics[scale=0.55]{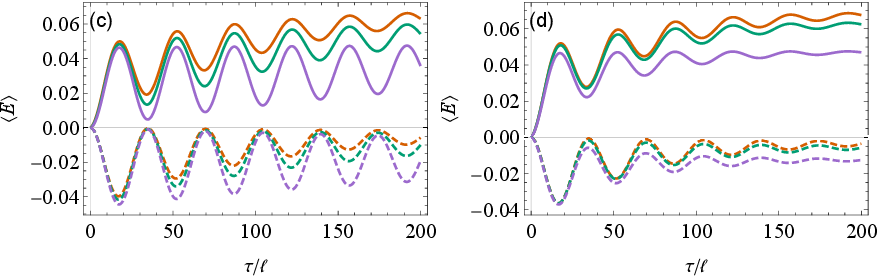}
 	\caption{ Evolution of average energy $ \langle E \rangle $ over time. We take $ \Delta \ell $ = 0.05, $ A \ell $ = 0.1 in (a)-(b) and  $ \Delta \ell $ = 0.15, $ A \ell $ = 0.1 in (c)-(d). Solid lines are associated with the quantum battery, and dashed lines are associated with the charger. Other parameters are the same as those in the graphs above.}
 	\label{fig6}
 \end{figure}

 From the above analysis, we find that whether the open quantum battery outperforms the closed one depends on the ratio of energy level spacing $\Delta \ell$ to charging amplitude $A \ell$. To better understand this charging performance physics, we investigate the flow of average energy during the charging process, as shown in Fig.~\ref{fig6}. It is found that when $\Delta \ell< A \ell$, curves associated with the quantum battery and the charger are not specular with respect to the $ x $-axis, this reflects that a certain amount of energy is dissipated into the reservoir. Similarly,  the curves associated with the quantum battery and the charger are also not specular with respect to the $ x $-axis when $\Delta \ell> A \ell$.
And the average energy $ \langle E_{\mathrm{B}}(\tau) \rangle $ change associated with the quantum battery is greater than the average energy $ \langle E_{\mathrm{C}}(\tau) \rangle $ change associated with the charger, which implies the fact that the average energy of the quantum battery is not only obtained from the charger but also extracted from the vacuum fluctuations of the quantum fields in rotating BTZ black hole, this result is consistent with the conclusion presented above.

In particular, from the dynamics of the average energy evolution, it is evident that the dissipation plays a dominant role in the overall charging process, as shown in Fig.~\ref{fig6}(c) and (d).
Notably, when the angular momentum $J/Ml = 0$, i.e., for a stationary BTZ black hole, the quantum battery extracts maximal energy from vacuum fluctuations via dissipative processes. However, as the angular momentum $J/Ml$ increases, i.e., as the BTZ black hole begins to spin, the energy the quantum battery extracts from vacuum fluctuations gradually decreases.
This means that the rotation of the BTZ black hole inhibits the quantum battery from extracting energy from the vacuum fluctuations.
  
 \noindent

\section{Conclusions}\label{Sec.4}
  
  In summary, we studied the dissipative dynamics of a quantum battery coupled to the massless scalar field of a rotating BTZ black hole. The effect of black hole rotation on the charging performance is explored, focusing on decoherence coupling regime. We find that the boundary conditions and the ratio between energy level spacing and charging amplitude determine how black hole rotation impacts the charging performance. When energy level spacing is smaller/larger than the charging amplitude, black hole rotation can enhance/weaken performance. Increasing rotational velocity strengthens the influence of boundary conditions on average energy dynamics. Moreover, energy extraction from vacuum fluctuations via dissipation in curved spacetime is demonstrated, with black hole rotation reducing extractable energy. In specific cases, such as extremal black holes, rotation can significantly enhance/weaken the charging performance. This reveals that quantum batteries in strong gravitational fields can probe critical black hole properties. Our study elucidates a new mechanism where black hole spin regulates quantum energy storage through vacuum fluctuations.
  
  	~~

	\acknowledgments
	
	This work was supported by the National Natural Science Foundation of China under Grants No.12475051, No.12374408, and No.12035005; the science and technology innovation Program of Hunan Province under grant No.2024RC1050; the Natural Science Foundation of Hunan Province under grant No.2023JJ30384; and the innovative research group of Hunan Province under Grant No.2024JJ1006. ZT was supported by the scientific research start-up funds of Hangzhou Normal University: 4245C50224204016.

	~~
	\appendix

\section{Charging performance indicators}\label{AppendixA}

 To describe the charging dynamics of quantum battery, the time evolution of spin component $\langle\sigma_m(\tau)\rangle$ is defined  as
\begin{equation}
	\langle\sigma_m(\tau)\rangle=\mathrm{Tr}[\rho(\tau)\sigma_m],
\end{equation}
where $\rho(\tau)=\mathrm{Tr}_\mathrm{F}[\rho_\mathrm{tot}(\tau)].$

We investigate the dynamical evolution of different subparts using the unitary rotation of the spin space

\begin{equation}\label{A2}
	\mathcal{R}=e^{-i\frac{\Theta}{2}\sigma_y}, 
\end{equation}
where the phase factor $ \Theta $ is selected as the value to project $ H_{\mathrm{B}} $ + $ H_{\mathrm{C}} $ only along the $ z $-axis. Following a unitary rotation, one obtains
\begin{equation}
	\begin{aligned}
		&
		\tilde{H}_{\mathrm{B}}=\mathcal{R}H_{\mathrm{B}}\mathcal{R}^{\dagger}=\frac{\Delta}{2}(\cos\Theta\sigma_{z}+\sin\Theta\sigma_{x}),\\
		&	\tilde{H}_{\mathrm{C}}=\mathcal{R}H_{\mathrm{C}}\mathcal{R}^{\dagger}=\frac{A}{2}(\sin\Theta\sigma_{z}-\cos\Theta\sigma_{x}),
		\\&
		\tilde{H}_{\mathrm{I}}=\mathcal{R}H_{\mathrm{I}}\mathcal{R}^{\dagger}=\lambda(\sin(\theta-\Theta)\sigma_{z}+\cos(\theta-\Theta)\sigma_{x})\otimes\phi(x(\tau)).
	\end{aligned}\label{eq11}
\end{equation}

 Moreover, by combining $\tilde{H}_{\mathrm{B}}$ and $\tilde{H}_{\mathrm{C}}$ in Eq. (\ref{eq11}), we can calculate both the average energy $\langle H_{\mathrm{B}}(\tau)\rangle$ of the quantum battery and the average energy $\langle H_{\mathrm{C}}(\tau)\rangle$ of the charger expressed in Eq. (\ref{eq17})
\begin{equation}
	\begin{aligned}
		&
		\langle H_{\mathrm{B}}(\tau)\rangle=\frac{\Delta}{2}\left[a_{3}(\tau)\cos\Theta+a_{1}(\tau)\sin\Theta\right], \\&
		\langle H_{\mathrm{C}}(\tau)\rangle=\frac{A}{2}\left[a_{3}(\tau)\sin\Theta-a_{1}(\tau)\cos\Theta\right].
	\end{aligned}
\end{equation}
The foregoing equations clearly show that the charge dynamics depend on the spacetime nature (\ref{eq16}), and in turn the charging dynamics can be used to learn about the spacetime nature. 

The charging dynamics of the quantum battery can be then determined by studying the energy changes between the different subparts at time $ \tau $
\begin{equation}
	\langle E_{j}(\tau)\rangle=\langle H_{j}(\tau)\rangle-\langle H_{j}(0)\rangle,
\end{equation}
where $ j $ = B, C, FI. For $\tau>0 $, the driving is assumed to be static, implying $\dot{H}=0$ and the system's energy balance 
$ 	\langle E_{\mathrm{B}}(\tau)\rangle+\langle E_{\mathrm{C}}(\tau)\rangle+\langle E_{\mathrm{FI}}(\tau)\rangle=0. $
The formula applies to arbitrary drive amplitude and coupling strength.

	\bibliographystyle{apsrev4-1}
	%

\end{document}